\documentclass{webofc}
\usepackage[varg]{txfonts}   
\wocname{European Physical Journal Web of Conferences}
\def\bb    #1{\hbox{\boldmath${#1}$}}

\woctitle{International Conference on New Frontiers in Physics 2013}
\begin{document}
\title{Foundation of Hydrodynamics of  Strongly Interacting Systems} 
%
%

\author{Cheuk-Yin Wong\inst{1}\fnsep\thanks{\email{wongc@ornl.gov}} }
\institute{Physics Division, Oak Ridge National laboratory, Oak Ridge,
  TN 37831, U.S.A.  }

\abstract
{
Hydrodynamics and quantum mechanics have many elements in common, as
the density field and velocity fields are common variables that can be
constructed in both descriptions.  Starting with the Schr\"odinger
equation and the Klein-Gordon for a single particle in hydrodynamical
form, we examine the basic assumptions under which a quantum system of
particles interacting through their mean fields can be described by
hydrodynamics.  
}  
\maketitle
\section{Introduction}
\label{intro}
 
\hspace*{0.4cm} 
Hydrodynamics has applications in many areas of
physics, for both finite and infinite systems
\cite{Boh39}-\cite{hydro}. We are interested in hydrodynamics of
quantum systems in which particles obey quantum mechanics, as for
example, in
\begin{itemize}
\item
an atomic nucleus as a liquid drop,

\item
a finite quark-gluon plasma,

\item
an assembly of hadrons at the end point of a nuclear collision,

\item
a droplet of Bose-Einstein condensate,

\item
an astrophysical object of nuclear matter or neutron matter.
\end{itemize}

It should be realized on the outset that hydrodynamics and quantum
mechanics have many elements in common.  For example, the density
field, $n(\bb r,t)$, and the velocity field, $\bb u(\bb r,t)$, are
common variables that can be constructed in both descriptions.  There
are correspondingly similar equations of motion for $n(\bb r,t)$ and
$\bb u(\bb r,t)$.  They also have elements that are different, as for
example in the relation between the pressure and the density in
classical hydrodynamics or between the pressure and the wave function
amplitude in quantum mechanics.  It is useful to compare and contrast
the similarities and differences so that their properties can be
better understood.

In the case of a nucleus with a large number of nucleons, the gross
static property can be described by the classical liquid drop model
\cite{Boh39,Hil53}.  There are however finite-size quantum shell
effects that arise from the quantization of single-particle states and
these quantum effects exert great influences on the static nuclear geometrical
configurations at their local energy minima.  The interplay of both
the classical bulk liquid-drop behavior and the quantum
single-particle effect has led to rich phenomena of many local
geometrical configurations built on top of a general underlying
liquid-drop background ("Funny Hills" as described in \cite{Bra72}).

A many-particle quantum mechanical system consists of many particles
which interact with other particles.  Much of the dynamics can be
understood by the the simple picture of single particles moving in the
mean-field $V(\bb r,t)$ generated by the other particles.  Quantum
effects will be pronounced in a finite system that is characterized by
discrete states with wave functions within a finite geometrical
region.  In dynamics involving such a finite system, as in the
evolution of the dense overlap region in high-energy nuclear collisions, the
finite quantum effects may be interest.  It is therefore useful to
examine to what extent quantum mechanics for the motion of a single
particle may be a part of the foundation for a hydrodynamical
description of a quantum many-body system.

\section{Schr\"odinger Equation in Hydrodynamical Form}
\label{sec-1}
 
\hspace*{0.4cm} 
To examine the behavior of one of the particles, we
can investigate the particle moving in the mean-field $V(\bb r,t)$
with the time-dependent Schr\"odinger equation
\begin{eqnarray}
i \hbar \frac{\partial }{\partial t} \psi (\bb r,t)  = - \frac{\hbar^2}{2m} \nabla ^2 \psi (\bb r,t) + V(\bb r,t)  \psi (\bb r,t). 
\end{eqnarray}
We follow Madelung \cite{Mad26} and Bohm \cite{Boh52} to write the
wave function in the form
\begin{eqnarray}
\psi (\bb r,t) = \phi (\bb r,t)\exp\{ i S(\bb r,t) - i \Omega (t)\},
\end{eqnarray}
where $\phi (\bb r,t)$, $S(\bb r,t)$, and $\Omega (t)$ are real
functions.  We construct $\psi^* i \hbar \partial_t \psi - \psi [i
  \hbar \partial_t \psi]^*$ and we get
\begin{eqnarray}
 \partial_t \phi^2 +     \nabla \cdot ( \phi^2
\frac{\hbar \nabla S}{m} )=0.
\end{eqnarray}
Upon identifying $\phi^2$ as the density field, $\rho$, and $\hbar
\nabla S/m$ as the velocity field, $ \bb u $, the above is just is the
equation of continuity for $\rho$ and $\bb u$.  We construct next
$\psi^* i \hbar \partial_t \psi + \psi [i \hbar \partial_t \psi]^*$
and we get
\begin{eqnarray}
\hbar \phi^2 (-2\partial_t S-4\partial_t \Omega) = \left \{ - \frac{\hbar^2}{2m} 2 [\phi \nabla^2\phi - 2 \phi^2 (\nabla S)^2 ] \right \}+ 2\phi ^2 V.
\end{eqnarray}
Dividing the above by $\phi^2$, taking the gradient $\nabla_i$, and
multiplying by $\phi^2$, we obtain
\begin{eqnarray}
\hbar \partial_t [ \phi^2 \nabla_i S] + \sum_{j=1}^3 \nabla_j\left  [ \frac{\hbar^2 \phi^2}{m} \nabla_iS \nabla_j S \right ] = -\frac{\hbar^2}{2m} \phi^2 \nabla_i \frac{\nabla^2 \phi}{\phi} - \phi^2 \nabla_i V.
\end{eqnarray}
If we define the quantum stress tensor as \cite{Won76}
\begin{eqnarray}
(i)~~~&& p_{ij}^{(q)}= -\frac{\hbar^2}{2m} \phi \nabla_i \nabla_j \phi +\frac{\hbar^2}{2m}\nabla_i  \phi \nabla_j \phi, 
\end{eqnarray}
then the time-dependent Schr\"odinger equation can be written in the
form
\begin{eqnarray}
\hbar\partial_t [ \phi^2 \nabla_i S] + \sum_{j=1}^3 \nabla_j\left  [ \frac{\hbar^2 \phi^2}{m} \nabla_iS \nabla_j S \right ]  = - \sum_{j=1}^3 \nabla_j p_{ij}^{(q)} -\phi^2 V.
\label{6}
\end{eqnarray}
This is just is the Schr\"odinger equation in hydrodynamical form,  
\begin{eqnarray}
 \partial_t [ \rho u_i ] + \sum_{j=1}^3 \nabla_j  [ \rho u_i u_j  ]
= - \frac{1}{m} \sum_{j=1}^3 \nabla_j p_{ij}^{(q)} -\frac{\rho}{m} V.
\label{7}
\end{eqnarray}
  The quantum stress tensor can also be written in two other
  alternative forms \cite{Won76},
\begin{eqnarray}
(ii)~~~&&p_{ij}^{(q)}=- \frac{\hbar^2}{4m}\delta_{ij}   \nabla^2 \phi^2 + \frac{\hbar^2}{m}\nabla_i \phi  \nabla_j \phi,
\\
{\rm or}~~~~~~(iii)~~~&&p_{ij}^{(q)}= +\frac{\hbar^2}{4m}\delta_{ij}   \nabla^2 \phi^2 - \frac{\hbar^2}{m}\phi \nabla_i \nabla_j \phi .
\end{eqnarray}

\section{An example of the quantum stress tensor}

We consider a plain wave with wave function $\phi(x)=A \cos (kx+B)$.
The quantum pressure is
\begin{eqnarray}
p_{xx}=A^2  \frac{\hbar^2}{2m} k^2,
\label{11}
\end{eqnarray}
which is a constant inside a spatially flat region, perpendicular to
the direction of motion.  It increases with the square of the wave
number, $k^2$, and is proportional to the kinetic energy of the
single-particle state.

We consider next an exponentially decaying wave function,
$
\phi(x)=B e^{-\kappa x}.
$ The quantum pressure is
\begin{eqnarray}
p_{xx}= -A^2  \frac{\hbar^2}{2m}\kappa^2 +A^2  \frac{\hbar^2}{2m}\kappa^2 =0.
\label{12}
\end{eqnarray}
The  stress tensor is zero beyond the point of decay.

We can now examine a square well of the form 
$V(x)=V_0 \Theta(a-|x|)$ with a geometrical width $a$.
A symmetric single-particle state wave functions is
\begin{eqnarray}
\psi(x)=A  \cos k x ~ \Theta (a-|x|) + B e^{-\kappa(| x|-a)}~\Theta (|x|-a);~~~
k=\sqrt{2 \mu (E-V_0)}; ~~~
\kappa = \sqrt{2 \mu E},
\end{eqnarray}
where $k$ and $\kappa$ satisfies the eigenvalue equation, $k \tan ka=
\kappa$.  An antisymmetric wave function can be similarly written down
by replacing $\cos kx$ by $\sin kx$.

\vspace*{0.1cm}

\begin{figure}[h]
 \hspace*{0.2cm}
\includegraphics[width=6.2cm]{sqrwell}
\label{fig-1}       
\end{figure}

 \protect\hangindent=-7.7cm \hangafter=0
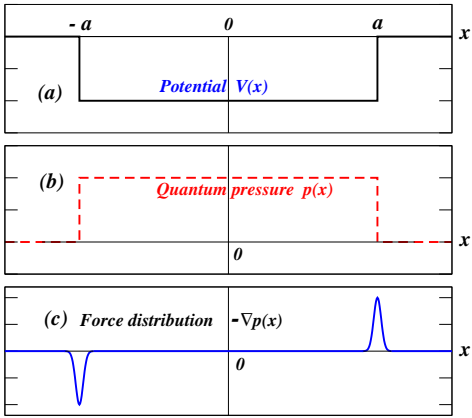
\noindent Fig. 1. The~single-particle~potential,~quantum
 \protect\linebreak  \hspace*{0.2cm} pressure, and force density 
in a square well.

\vspace*{-6.5cm} 
\hangindent=6.9cm 
\hangafter=0 
For a wave function with $n$ numbers of nodes in a very deep potential
well, the quantum pressure is
\begin{eqnarray}
p_{xx}= A^2  \frac{\hbar^2}{2m} k^2 \sim  
 \frac{\hbar^2}{ma^3}\frac{(2n+1)^2\pi^2}{8},
\end{eqnarray}
which reveals that the quantum stress tensor $p_{xx}(x)$ is constant
inside the square well, proportional to the kinetic energy of the
single particle measured relative to the bottom of the well.  It
depends sensitively on the geometrical width $a$ of the finite well,
in contrast to a particle in an infinite medium without a boundary.
The quantum stress tensor is zero outside the square well as depicted
in Fig.\ 1 (b).  As a consequence, the force distribution $-\nabla
p(x)$ is sharply peak at the boundary and zero in the interior as in
Fig.\ 1(c), directing outward.

In a self-assembled system with many single particles, the mean-field
potential can be approximately represented by a square well.  If the
mean-field potential is suddenly weakened by the occurrence of
a phase transition that alters the nature of the constituent
interactions in the new phase,  or if the external confining potential 
is suddenly removed,  then hydrostatic equilibrium cannot be maintained and a
hydrodynamical flow of a pressurized medium will then occur.

\section{Hydrodynamical Description of a Nucleus }

We can examine a many-body system with strongly interacting
constituents in the extended mean-field approximation
\cite{Won76,Won77,Ayi80,Lac99} which is represented by a set of
single-particle states $\psi_a$ in their mean field potential and a
set of occupation numbers $n_a$.

The Schr\"odinger equation for the single-particle wave function
$\psi_a$ in the mean field $V(\bb r,t)$ is
\begin{eqnarray}
i \hbar \frac{\partial }{\partial t} \psi_a (\bb r,t)  = - \frac{\hbar^2}{2m} \nabla ^2 \psi_a (\bb r,t) + V(\bb r,t)  \psi_a (\bb r,t), 
\end{eqnarray}
and the occupation number $n_a$ obeys the master equation
\begin{eqnarray}
\hspace*{-0.8cm}
\frac{\partial n_a }{\partial t} = \sum_{234}\frac{2\pi}{\hbar}|\langle a2|v'|34 \mp 43\rangle|^2
D(\epsilon_a + \epsilon_2-\epsilon_3 + \epsilon_4)
[ n_a n_2 (1\mp n_3) (1 \mp n_4) - n_3 n_4 (1 \mp n_a) (1 \mp n_2)],
\end{eqnarray}
where $D(x)$ is a delta-function type distribution with a width, the
upper sign is for fermions and the lower sign is for bosons, $v'$ is
the residual interaction, and $\epsilon_a$ is the expectation value of
the single-particle Hamiltonian for the state $\psi_a$ that can be a
slow function of time.

As the Schr\"odinger equation for a single-particle in an external
field can be cast into a hydrodynamical form, we examine under what
conditions a system of single particles interacting in their own mean
fields can also be cast into a hydrodynamical form.  We write
\begin{eqnarray}
\label{decom}
\psi_a ({\bf r},t)=\phi_a({\bf r},t) \exp\{-i S_a({\bf r},t) - i \Omega_a (t) \}.
\end{eqnarray}
We get
\begin{eqnarray}
 \partial_t \phi_a^2 +     \nabla \cdot ( \phi_a^2
\frac{\hbar \nabla S_a}{m} )=0.
\end{eqnarray}
We multiple by $n_a$ and we get 
\begin{eqnarray}
 \partial_t [ n_a \phi_a^2]  + n_a  \nabla \cdot \left [  \phi_a^2 
\frac{\hbar \nabla S_a}{m} \right ]=  \phi_a^2 [\partial_t n_a].
\end{eqnarray}
Sum over $a$, we get,
\begin{eqnarray}
 \partial_t  \biggl [\sum_a n_a \phi_a^2  \biggr ]
+     \nabla \cdot  \biggl [\sum_a n_a  \phi_a^2
\frac{\hbar \nabla S_a}{m}  \biggr ]=- \sum_a   \phi_a^2 [\partial_t n_a].
\label{19}
\end{eqnarray}
We can introduce the density matrix
\begin{eqnarray}
{\cal N} (\bb r, \bb r';  t) = \sum_a n_a(t) \psi_a^*  (\bb r,t)  \psi_a (\bb r',t).
\end{eqnarray}
The diagonal element of the density matrix is the density field
density field $\rho(\bb r,t)$
\begin{eqnarray}
\rho (\bb r, t) = \sum_a n_a\psi_a^*  (\bb r,t)  \psi_a (\bb r,t).
\end{eqnarray}
We can also introduce the velocity field $\bb u(\bb r, t)$
\begin{eqnarray}
\rho (\bb r, t)  \bb u(\bb r, t) = \sum_a n_a\phi_a^2  (\bb r,t)  \frac{\hbar \nabla S_a (\bb r,t)}{m}.
\end{eqnarray}
Equation (\ref{19}) is then the equation of continuity
\begin{eqnarray}
\!\!\!\!\!\!\!\!\!\!\partial_t \rho(\bb r,t)
+    \nabla \cdot  [ \rho (\bb r,t) ~\bb u (\bb r,t)] 
\!\!\!\!\!&=&\!\!\!\!\! - \sum_a   \phi_a^2 [\partial_t n_a]
\label{24}
\\
\!\!\!\!\!&=&\!\!\!\!\!
\frac{1}{4} \sum_{1 234} [\phi_1^2  +
\phi_2^2-\phi_3^2-\phi_4^2]
\frac{2\pi}{\hbar}|\langle a2|v'|34\mp 43\rangle|^2
\nonumber\\
\!\!&  &\!\!\!\!\times \,
D(\epsilon_a + \epsilon_2-\epsilon_3 + \epsilon_4)
[ n_a n_2 (1+\mp n_3) (1 \mp n_4) - n_3 n_4 (1\mp n_a) (1\mp n_2)] ,\nonumber
\end{eqnarray}
which contains an additional term on the right-hand side relating the
scattering of a pair of particles with wave functions $\phi_1$ and
$\phi_2$ non-locally into states of $\phi_3$ and $\phi_4$.  Here, the
factors $[\phi_1^2 + \phi_2^2-\phi_3^2-\phi_4^2]$ and $[ n_a n_2 (1\mp
  n_3) (1\mp n_4) - n_3 n_4 (1\mp n_a) (1 \mp n_2)]$ are antisymmetric
in the exchange of $12$ with $34$, so the above sum is not zero.
However, the spatial integral of the right-hand side is zero.  So, the
total number of particles are conserved, even though a pair of
particles can scatter into another pair non-locally.

For the single particle in state $a$, we can get the equation of
motion for the probability current $ \phi_a^2 \nabla_i S_a$
\cite{Won77},
\begin{eqnarray}
\partial_t [ \phi_a^2 \nabla_i S_a] + \sum_{j=1}^3 \nabla_j\left  [ \frac{\phi^2}{m} \nabla_iS_a \nabla_j S_a \right ]  = - \sum_{j=1}^3 \nabla_j p_{ij}^{(q)}(a) -\phi_a^2 V.
\label{6}
\end{eqnarray}
We multiple by $n_a$ and sum over $a$, and we get the Euler equation
for $n u_i$
\begin{eqnarray}
\frac {\partial n u_i}{\partial t} + \sum_{j=1}^3
n u_i u_j =  -\frac{1}{m} \sum_{j=1}^3 
\nabla_j \left ( p_{ij}^{(q)}+p_{ij}^{(t)}+p_{ij}^{(v)}\right )
- \sum_a   \phi_a^2\nabla_i S_a [\partial_t n_a].
\label{25}
\end{eqnarray}
where  
\begin{eqnarray}
p_{ij}^{(q)}=-\frac{\hbar^2}{4m} \sum_a n_a \nabla^2 \phi_a^2 \delta_{ij}
+\frac{\hbar^2}{m} \sum_a n_a \nabla_i \phi_a \nabla_j \phi_a,
\label{27}
\end{eqnarray}
\begin{eqnarray}
p_{ij}^{(t)}=\frac{\hbar^2}{m} \sum_a n_a 
\phi_a^2 (\nabla_i S_a - m u^i) (\nabla_j S_a - m u^j),
\end{eqnarray}
\begin{eqnarray}
\label{non}
\nabla_j p_{ij}^{(v)}({\bf r},t)=
\int d^3 {\bf r}_2[ \rho({\bf r},t) \rho({\bf r}_2,t) 
\mp {\cal N} (\bb r,\bb r_2; t) {\cal N} (\bb r_2,\bb r; t) ] 
{\nabla_j}v({\bf r},{\bf r}_2),
\label{29}
\end{eqnarray}
and $v({\bf r},{\bf r}_2)$ is the two-body interaction that generates
the mean field.  The last term in Eq.\ (\ref{25}) arises from the
change in current in the scattering of a pair of particles from state
12 to state 34 due to residual interactions $v'$.

The equation of continuity (\ref{24}) and the Euler equation
(\ref{25}) contain terms that depends on the change of the occupation
probability $[\partial_t n_a]$.  A finite system for which $n_a$ is
quasi-stationary is said to be in thermal equilibrium, which can be
attained when $\partial_t n_a$$\sim$ 0.  At thermal equilibrium, the occupation numbers obey  
\begin{eqnarray}
n_a = \frac{1}{1\mp \exp \{ (\epsilon_a - \mu)/T\}}, 
\end{eqnarray}
characterized by a Fermi energy $\mu$ and a temperature $T$ that can
be a slow function of time. Here in the above equation, the upper sign is for fermions and the lwoer sign for boson.  If a system with an initial occupation
number $n_a$ that is far from the thermal equilibrium distribution, it
will eventually reach thermal equilibrium after a thermal
equilibriation time.  When thermal equilibrium is reached,
$[\partial_t n_a]=0$, and Eqs.\ (\ref{24}) and (\ref{25}) become the
equations of motion in hydrodynamics.  The requirement of thermal
equilibrium is one of the basis for hydrodynamics.

After thermal equilibrium is reached, the total hydrodynamical
pressure arises from many sources as indicated in
Eqs.\ (\ref{25})-(\ref{29}) : (i) the mean-field stress tensor
$p_{ij}^{(v)}$ from the mutual interaction between fluid elements,
(ii) the quantum stress tensor $p_{ij}^{(q)}$ from quantum effects,
and (iii) the thermal stress tensor $p_{ij}^{(t)}$ from the deviation
of the individual velocity fields from the local mean velocities.  The
stress tensor due to the mean-field interaction can also be given as
\begin{eqnarray}
\frac{\partial}{\partial x^j} p_{ij}^{(v)}({\bf r},t)=
\rho ({\bf r},t) \frac{\partial}{\partial x^j}
\left ( \frac{\partial W^{(v)}( \rho ) }{\partial \rho } \right ),
\end{eqnarray}
where $W^{(v)}$ is the energy per particle arising from the mean-field
interaction.  The mean-field stress tensor $ p_{ij}^{(v)}$ is given
explicitly by
\begin{eqnarray}
\label{pw}
 p_{ij}^{(v)}=
\left \{\rho  \frac{\partial (W^{(v)} \rho) }{\partial \rho}
- W^{(v)} \rho \right \}\delta_{ij}.
\end{eqnarray}
For example, for a nucleus in which the nucleons interact with the
Skyrm interaction, we have \cite{Won77}
\begin{eqnarray}
 p_{ij}^{(v)}=\frac{3}{8} ( t_0 +\frac{1}{3} t_3 \rho  ) \rho^2  \delta_{ij}, 
\end{eqnarray}
where for the Skyrm I nucleon-nucleon interaction, $t_0=-1057$
MeV/fm$^3$ is the two-body interaction strength, and $t_0=+14463$
MeV/fm$^6$ is the three-body interaction strength.

The quantum stress tensor of a finite system with discrete states
depends on the geometry of the system.  Its magnitude varies inversely
with the corresponding width of the system in that direction and needs
to be examined on a case-by-case basis.

For a large fermion system with the neglect of the geometrical
dependence, one can consider the Thomas-Fermi approximation of
treating the single-particle states as a continuum and the quantum
stress tensor for fermions is
\begin{eqnarray}
 p_{ij}^{(q)}=
\frac{\hbar^2}{5 m}
\left ( \frac{6 \pi^2}{4} \right )^{2/3} \rho^{5/3} \delta_{ij}. 
\end{eqnarray}
A nucleus is a strongly-coupled system.  The quantum and mean-field
stress tensors are the dominant component for the nuclear fluid at low
and moderate temperatures.  The thermal stress tensor $p_{ij}^{(t)}$
can take on different values, depending on the occupation numbers of
the single-particle states that determines the degree of thermal
equilibrium of the system.  For a thermally equilibrated fermion
system, the thermal stress tensor in the approximation of treating it
as a local fermion gas is
\begin{eqnarray}
 p_{ij}^{(t)}=
\frac{\hbar^2}{5 m}
\left ( \frac{6 \pi^2}{4} \right )^{2/3} \rho^{5/3} 
\left [ \frac{2mkT}{\hbar^2 m (6 \pi^2 \rho/4)^{2/3}} \right ]^2 \delta_{ij},
\end{eqnarray}
which is small for low and moderate temperatures.  Thus, when
$|p_{ij}^{(q)}+p_{ij}^{(v)}| \gg p_{ij}^{(t)}$ in a strongly
interacting system at low and moderate temperatures, there can be
situations in which the system can behave quasi-hydrodynamically even
though the state of the system has not yet reach thermal equilibrium,
as is evidenced by the presence of nuclear collective vibrational and
rotational states at low and moderate temperatures.  In this case, the
hydrodynamical state is maintained essentially by the quantum stress
tensor and the strong mean-field stress tensor, and not by the thermal
stress tensor.

\section{ Klein-Gordon equation in hydrodynamical form}

We wish to examine next a finite system of relativistic bosons as an
assembly of single-particles obeying the Klein-Gordon equation and
interacting with the mean-field generated by other particles.  We
would like to write the Klein-Gordon equation in hydrodynamical form
in terms of the density and velocity fields. Here one encounters the
well-known problem in Klein-Gordon equation that the naive probability
density $\rho$=$2 {\rm Im} (\psi^* \partial_t \psi)$ constructed from
the wave function $\psi$ is not necessarily a positive quantity.  The
presence of a negative probability density may appear to preclude its
description in relativistic hydrodynamics.

A resolution of the pathology was provided by Feshbach and Villars
\cite{Fes58} who reintroduced wave mechanical interpretation of the
wave field $\psi$ by noting that the Klein-Gordon equation actually
contains both particle and antiparticle degrees of freedom.  The
Klein-Gordon equation can be cast as a set of coupled time-dependent
Schr\"odinger equations for the particle and antiparticle wave
function components both with positive probability densities
\cite{Fes58}.

We shall follow Feshbach-Villar's method and consider a
single-particle with a charge $e$ and a rest mass $m_0$ in a scalar
field $\cal S$ and a gauge field $A_\mu$ \cite{Won10}.  The
Klein-Gordon equation for the wave function of the single-particle is
\begin{eqnarray}
\biggl  \{ (i\hbar \partial_t  - e A_0 )^2 -(-i\hbar \nabla -  e \bb A)^2  - (m_0+{\cal S})^2 \biggr \} \Psi (r,t) = 0,
\label{kg}
\end{eqnarray}
where for brevity of notation, we shall abbreviate $m_0+{\cal S}$ by
$M$.  We consider the initial energy of the system to be $E$ which
includes the rest mass $m_0$ so that $E$ is positive definite.  If the
external interactions are time-independent, then $E$ is a constant of
motion.  If the external fields are time-dependent, then $E$ changes
with time and we need $E$ and $\partial_t E$ initially to start the
time evolution.  We can get $E>0$ as the expectation value of $i \hbar
\partial_t$ and obtain $\partial_t E$ by an iterative procedure.

We  introduce an auxiliary wave function $\Psi_4$,
\begin{eqnarray}
(i\hbar \partial_t - e A_0)  \Psi = (E -  e A_0) \Psi_4.
\label{psi}
\end{eqnarray}
Then the Klein-Gordon equation (\ref{kg}) becomes
\begin{eqnarray}
(i\hbar \partial_t- e A_0)^2\Psi = \Psi_4 [ i\hbar \partial_t (E- e A_0)]
+(E - e A_0)(i\hbar \partial_t- e A_0)\Psi_4,
\end{eqnarray}
which allows us to obtain the equation for $(i\hbar \partial_t- e A_0)\Psi_4$ as
\begin{eqnarray}
(i\hbar \partial_t- e A_0)\Psi_4
=\frac{1}{(E- e A_0)}\biggl  \{
\{ (-i\hbar \nabla- e \bb A)^2  + M^2 \}\Psi
 -  \Psi_4 [ i\hbar \partial_t (E- e A_0)]\biggr \}.
\label{psi4}
\end{eqnarray}
By the method of Feshbach and Villar \cite{Fes58}, the Klein-Gordon
equation (\ref{kg}) that is second order in time for $\Psi$ becomes a
set of coupled equations (\ref{psi}) and (\ref{psi4}) that are first
order in time for $\Psi$ and $\Psi_4$.  What remains is to turn the
wave amplitudes $\Psi$ and $\Psi_4$ into particle and antiparticle
wave function components.  This can be accomplished by forming the sum
and difference of Eqs. (\ref{psi}) and (\ref{psi4}),
\begin{eqnarray}
\!\!\!\!\!\!\!\!\!\!\!\!\!\!\!\!(i\hbar \partial_t - e A_0)  [ \Psi + \Psi_4]  =
\frac{1}{(E- e A_0)}\biggl  \{
[ (-i\hbar \nabla-e \bb A)^2  + M^2 ]\Psi
 \!+\!(E- e A_0)^2 \Psi_4 \!-\!  \Psi_4 [i\hbar \partial_t (E- e A_0)]\biggr \},
\\
\!\!\!\!\!\!\!\!\!\!\!\!\!\!\!\!(i\hbar \partial_t- e A_0) [ \Psi - \Psi_4]  =
\frac{1}{(E- e A_0)}\biggl  \{
- [ (-i\hbar \nabla-e \bb A)^2 \! + \! M^2 ]\Psi
 \!+\!(E- e A_0)^2 \Psi_4 \!+\!  \Psi_4 [ i\hbar \partial_t (E- e A_0)]\biggr \}.
\end{eqnarray}
We identify the sum $\Psi$+$\Psi_4$ as the particle component $\psi_e$
of the wave function, and the difference $\Psi$-$\Psi_4$ as the
antiparticle component $\psi_{\bar e}$ of the wave function by
defining
\begin{eqnarray}
\psi_e=\frac{\Psi + \Psi_4}{{2}} ~~~{\rm and~~}
\psi_{\bar e}^*=\frac{\Psi - \Psi_4}{{2}},
\end{eqnarray}
so that 
\begin{eqnarray}
\Psi={\psi_e + \psi_{\bar e}^*} ~~~{\rm and~~}
\Psi_4={\psi_e - \psi_{\bar e}^*}.
\end{eqnarray}
We then obtain a set of equations coupling the particle and
antiparticle components of the wave functions \cite{Won10},
\begin{eqnarray}
\!\!\!\!\!\!\!\!\!\!\!\!(i\hbar \partial_t- e A_0) \psi_e =
\frac{1}{2(E- e A_0)}\biggl  \{
\biggl [ (-i\hbar \nabla- e \bb A))^2  + M^2 
 +[(E- e A_0)^2 -i\hbar \partial_t (E- e A_0)] \biggr ]
 \psi_e
\nonumber\\
+ 
\biggl [ (-i\hbar \nabla- e \bb A))^2  + M^2
 -[(E- e A_0)^2 -i\hbar \partial_t (E- e A_0)]
\biggr ]\psi_{\bar e}^* \biggr \},
\label{43}
\\
\!\!\!\!\!\!\!\!\!\!\!\!(i\hbar \partial_t + e A_0)\psi_{\bar e} =
\frac{1}{2(E- e A_0)}\biggl  \{
\biggl [ (-i\hbar \nabla+e \bb A)^2  + M^2
 +[(E- e A_0)^2 -i\hbar \partial_t (E- e A_0)]
\biggr ]
\psi_{\bar e} 
\nonumber\\
 ~~~~~~~~~~~~~~~+ 
\biggl [ (-i\hbar \nabla+e \bb A)^2  + M^2 
 -[(E- e A_0)^2 -i\hbar \partial_t (E- e A_0)]
 \biggr ]
\psi_e^*
\biggr \}.
\label{44}
\end{eqnarray}
Thus, by the Feshbach-Villars method, the Klein-Gordon equation that
is second-order in the time derivative can be decomposed into a set of
coupled first-order Schr\"odinger equations containing particle wave
function component $\psi_e$ and antiparticle wave function component
$\psi_{\bar e}$.  In these coupled equations, $\psi_e$ and $\psi_{\bar
  e}$ have positive norms, $|\psi_e|^2$ and $|\psi_{\bar e}|^2$, which
can be interpreted as the densities of the probability fluid of
particles and antiparticles respectively, as in hydrodynamics.

For simplicity of notation, we shall denote $\psi_e$ by $\psi_+$ and
$\psi_{\bar e}$ by $\psi_-$.  A general solution of the Klein-Gordon
equation for a particle with a charge $e$ contains a predominant
particle component, $\psi_e$=$\psi_+$, and a small antiparticle
component, $\psi_{\bar e}$=$\psi_-$.  The above set of coupled
equations (\ref{43}) and (\ref{44}) for a particle with a charge $e$
and a positive energy $E$ can be re-written compactly as
\begin{eqnarray}
\!\!\!\!\!\!\!\!\!\!\!(i\hbar\partial_{t}\mp  eA^{0})\psi_\pm=\frac{1}{2(E-e A^0)}
\biggl \{
\biggl [ (\frac{\hbar}{i}\nabla\mp  e\bb {A})^{2}+M^{2}+[(E-e A^0)^{2}-i\hbar \partial_t(E-e A^0)] \biggl ]\psi_\pm
\nonumber\\
\,\,\,\,\,\,\,~~~~~~~~~~~~~~~~~~
+\biggl [  (\frac{\hbar}{i}\nabla\mp e\bb {A})^{2}
+M^{2}-[(E-e A^0)^{2}-i\hbar \partial_t(E-e A^0)]
\biggr ]
\psi_\mp^* \biggr \}.
\label{45}
\end{eqnarray}

Similar to the above, the wave function for an antiparticle with a
charge $\bar e$=$-e$ and a positive energy $E$ contains a predominant
antiparticle component $\psi_{\bar e}$(=$\psi_-$) and a small particle
component $\psi_{e}$(=$\psi_+$) with positive norms, $|\psi_\pm|^2$.
The corresponding set of coupled equations for such an antiparticle
can be obtained from the above Eq.\ (\ref{45}) by changing $e \to -e$
and $\pm \to \mp$ to yield
\begin{eqnarray}
\!\!\!\!\!\!\!\!\!\!\!(i\hbar\partial_{t}\pm  eA^{0})\psi_\mp=\frac{1}{2(E+e A^0)}
\biggl \{
\biggl [ (\frac{\hbar}{i}\nabla\pm  e\bb {A})^{2}+M^{2}+[(E+e A^0)^{2}-i\hbar \partial_t(E+e A^0)] \biggl ]\psi_\mp
\nonumber\\
\,\,\,\,\,\,\,~~~~~~~~~~~~~~~~~~
+\biggl [  (\frac{\hbar}{i}\nabla\pm e\bb {A})^{2}
+M^{2}-[(E+e A^0)^{2}-i\hbar \partial_t(E+e A^0)]
\biggr ]
\psi_\pm^* \biggr \}.
\label{46}
\end{eqnarray}

Equation (\ref{45}) for a predominantly particle state and
Eq. (\ref{46}) for a predominantly antiparticle state can be further
rewritten in a more succinct form by introducing $e_\pm=\pm e$ as
\cite{Won10}
\begin{eqnarray}
\!\!\!\!\!\!\!\!\!\!\!(i\hbar\partial_{t}-  e_\pm A^{0})\psi_\pm=\frac{1}{2(E-e_\pm A^0)}
\biggl \{
\biggl [ (\frac{\hbar}{i}\nabla - e_\pm \bb {A})^{2}+M^{2}+[(E-e_\pm A^0)^{2}-i\hbar \partial_t(E-e_\pm  A^0)] \biggl ]\psi_\pm
\nonumber\\
\,\,\,\,\,\,\,~~~~~~~~~~~~~~~~~~
+\biggl [  (\frac{\hbar}{i}\nabla - e_\pm \bb {A})^{2}
+M^{2}-[(E-e_\pm A^0)^{2}-i\hbar \partial_t(E-e_\pm  A^0)]
\biggr ]
\psi_\mp^* \biggr \},
\label{47}
\end{eqnarray}
with $e_+=e$ for a predominantly particle state with a large $\psi_+$
component, and $e_+=-e$ for a predominantly antiparticle state with a
large $\psi_-$ component.
 
The second term inside the curly bracket on the right-hand side of the
equation (\ref{47}) represents the particle-antiparticle coupling and
pair production.  It involves essentially the difference between of
$E^2$ and ${\bf p}^2+(m_0+{\cal S})^2$ that is quite small when the
strength of the interaction relative to the energy (or rest mass) of
the particle is small.

\section{Equations of continuity for fluids of particles and antiparticles}

To see how the probability fluids of a particle behave in space and
time, we consider the set of coupled equation (\ref{47}) for the two
components, $\psi_\pm$, in terms of their amplitudes and phase
functions,
\begin{eqnarray}
\psi_\pm({\bf r},t)=\phi_\pm ({\bf r},t) e^{iS_\pm({\bf r},t)-i\Omega_\pm (t)}.
\label{48}
\end{eqnarray}
We construct
$\psi_\pm^*$$\times$(\ref{47})-$\psi_\pm$$\times$(\ref{47})$^*$.
After some manipulations, we find
\begin{eqnarray}
\label{con}
\partial_{t}[(E-e_\pm A^0) \phi_\pm^2]
+ \nabla\cdot [\phi_\pm^2(\nabla S_\pm -e_\pm {\bf A}) ]
=X_\pm,
\end{eqnarray}
where
\vspace*{-0.9cm}
\begin{eqnarray}
&&2X_\pm =\{ \chi_\pm^*(\frac{\hbar}{i}\nabla-e_{\pm}{\bf A})^{2}\chi_\mp^*
- \chi_\pm(\frac{\hbar}{-i}\nabla-e_{\pm}{\bf A})^{2}\chi_\mp \}
\nonumber\\
&& ~~~~~~~~~~ +[(m_0+{\cal S})^2+(E-e_\pm A)^2] 
(\chi_\pm^*\chi_\mp^*-\chi_\pm \chi_\mp)
 +
[i\hbar \partial_t(E-e_\pm A^0)] (\chi_\pm^*\chi_\mp^*+\chi_\pm\chi_\mp).
\end{eqnarray}
The quantities $X_+$ and $X_-$ are not generally a full divergence.
The total number of particles and antiparticles in the two components
are not separately conserved due to the production of particle pairs.
However, the difference of the particle number and antiparticle
numbers of the two components satisfies the equation
\begin{eqnarray}
\partial_{t}[(E-e_\pm A^0) (\phi_+^2-\phi_-^2)]
+ \nabla\cdot [\phi_+^2(\nabla S_+ -e_+{\bf A})] 
- \nabla\cdot [\phi_-^2(\nabla S_- -e_- {\bf A})] 
=
X_+-X_-,
\label{51}
\end{eqnarray}
where $X_+$-$X_-$ is a complete divergence,
\begin{eqnarray}
\label{XX}
\!\!\!\!\!\!\!\!\!\!\!\!\!\!X_+-X_-
\!= \! -
\nabla \cdot (\psi_+^* \nabla \psi_-^*-\psi_-^* \nabla \psi_+^*)/2
+\nabla \cdot (\psi_+ \nabla \psi_- - \psi_- \nabla \psi_+)/2
-
\nabla \cdot [e_+ A^0 ( \psi_+^*\psi_-^*+\psi_+ \psi_-)/i].
\end{eqnarray}
Therefore, the quantity 
\begin{eqnarray}
\label{eq38}
n_{\rm particle}=\int d{\bf r} \frac {E-e_\pm A^0}{m_0} (\phi_+^2-\phi_-^2)
=\int d{\bf r} \frac {E-e_\pm A^0}{m_0} (|\psi_+|^2-|\psi_-|^2)
\end{eqnarray}
is a conserved quantity because the right-hand side of the equation
(\ref{XX}) is a complete divergence.  The additional number of
particles produced is equal to the additional number of antiparticles
produced.  The equal increase in particle and antiparticle numbers
associated with $X_+$ and $X_-$ represents the occurrence of
particle-antiparticle pair production by the mean field.  A single-particle solution
with a $n_{\rm particle}=n_\pm=\pm 1$ is one in which $|\psi_\pm|^2
\gg |\psi_\mp|^2$ and can be normalized to be
\begin{eqnarray}
\label{na}
\int d^3r \frac{E-e_\pm A^0}{m} [|\psi_\pm|^2 -|\psi_\mp|^2]=1 {\rm ~~ for~a~particle~state~with~}n_{\rm particle}=\pm 1.
 \end{eqnarray}

Next, to obtain the Euler equation, we construct
$\psi_\pm^*$$\times$(\ref{47})+$\psi_\pm$$\times$(\ref{47})$^*$, and
we get
\begin{eqnarray}
\label{euler}
&&
\!\!\!\!\!\!\!\!\!\!\!\!\!\!\!\!\!\!
\phi_\pm^2 (-2\partial_t S_\pm-2e_\pm A^0 ) 
\nonumber\\
&& =
\frac{1}{2(E-e_\pm A^0)} \biggl  \{
-[2\phi_\pm \nabla^2 \phi_\pm-2\phi_\pm^2(\nabla S_\pm
-e_\pm {\bf A} )^2 ]
+[(m_0+{\cal S})^2+(E-e_\pm A^0)^2] 2\phi_\pm^2 
\nonumber\\
&&~~+
\psi_\pm^*(\frac{\hbar}{i}\nabla-e_{\pm}{\bf A})^{2} \psi_\mp^*
+\psi_\pm(\frac{\hbar}{-i}\nabla-e_{\pm}{\bf A})^{2}\psi_\mp 
+
[(m_0+{\cal S})^2-(E-e_\pm A)^2] 
(\psi_\pm^*\psi_\mp^*+\psi_\pm \psi_\mp)
\nonumber\\
&&~~ +
[i\hbar \partial_t(E-e_\pm A^0)] (\psi_\pm^*\psi_\mp^*-\psi_\pm\psi_\mp)
\biggr \}.
\label{55}
\end{eqnarray}

The quantities on the right-hand sides of Eqs.\ (\ref{51}) and
Eq.\ (\ref{55}) contain terms of binary products such as $\psi_\pm^*
\psi_\pm^*$ and $\psi_\pm \psi_\pm$. According to Eq.\ (\ref{47}),
they contain time factors $e^{2i\Omega(t)}$ with a time frequency
greater than $2m_0/\hbar$ and they represent zitterbewegung motion of
the coupling between the particle and antiparticle density fields and
the current fields.  Zitterbewegung leads to pair production, but the
time average of these $e^{2i\Omega(t)}$ contributions over a long
period of time gives
\begin{eqnarray}
\langle \psi_\pm^* \psi_\pm^* + \psi_\pm \psi_\pm \rangle_T \sim \frac{\hbar}{2 m_0 T}.
\end{eqnarray}
If the dynamical time scale $T \gg {1}/{2m_0}$, then terms of the type $\psi_\pm^* \psi_\pm^*$
and $\psi_\pm \psi_\pm$ in $X_\pm$ becomes negligible  when averaged over the time period $T$.

A hydrodynamical description will be appropriate after the stage of
active pair production has passed and the expansion of the system is
now driven by a slowly varying external field (or a mean field), with
a dynamical time scale $T$ much greater than $\hbar /2m_0$.  It is
this type of motion for which we wish to provide a hydrodynamical
description.  In that case, the contributions from pair production and
zitterbewegung motion by the mean field averaged over the time scale for mean-field
motion may be small and neglected.  We get uncoupled equations of
motion for two kinds of particles.  This is equivalent to the case of
a ``simple fluid" in relativistic hydrodynamics, in which the chemical
composition of the fluid either ceases to change \cite{Lan59} or
changes according to the requirement of thermodynamic equilibrium.

\section{Euler equation in the approximation of no pair production}

Under the circumstance when the pair-production and zitterbewegung arising from the mean-field can
be neglected, terms of $\psi_\pm^* \psi_\pm^*$ and $\psi_\pm \psi_\pm$
in Eqs.\ (\ref{51}) and (\ref{55}) can be neglected.  After dividing
Eq.\ (\ref{55}) by $-2\phi_\pm^2$, the equation for the phase function
$S_\pm$ for this simplified case is
 \begin{eqnarray}
(\partial_t S_\pm+e_\pm A^0 ) 
=
\frac{1}{2(E-e_\pm A^0)} \biggl  \{
[(\nabla^2 \phi_\pm)/\phi_\pm-(\nabla S_\pm
-e_\pm {\bf A} )^2 ]
-(m_0+{\cal S})^2-(E-e_\pm A^0)^2  \biggl  \}.
\end{eqnarray}
For this case with suppressed pair production, $e_\pm=n_\pm e =
\pm e$.  We take the gradient $\nabla_i$ of the above for $i=1,2,3$,
and multiply by $\phi_\pm^2(E-e_\pm A^0)$.  We obtain
\begin{eqnarray}
&&\!\!\!\!\!\!\!\!\!\!\!\!\!\!\!\!
(E-e_\pm A^0)\phi_\pm^2 \partial_t (\nabla_i S_\pm-e_\pm A^i) 
\nonumber\\
&& = \biggl \{ \phi_\pm^2 \nabla_i
[(\nabla^2 \phi_\pm)/2\phi_\pm]
-  \sum_{j=1}^3 \phi_\pm^2 (\nabla_j S_\pm-e_\pm  A^j ) \nabla_j (\nabla_i S_\pm-e_\pm A^i ) 
\nonumber\\
&&~~~
-(m_0+{\cal S})\phi^2 \nabla_i {\cal S}-\sum_{j=1}^3 \phi_\pm^2 (\nabla_j S_\pm-e_\pm  A^j )e_\pm F^{ij}
\biggr \}- (E-e_\pm A^0)\phi_\pm^2e_\pm F^{0i}
\nonumber\\ 
&&~~~
+\frac{e_\pm \phi_\pm^2 \nabla_i  A^0}{2(E-e_\pm A^0)} \biggl  \{
(\nabla^2 \phi_\pm)/\phi_\pm-(\nabla S_\pm-e_\pm {\bf A} )^2 ]
-(m_0+{\cal S})^2+(E-e_\pm A^0)^2\biggr \}.
 \nonumber\\ 
\end{eqnarray}

Using the equation of continuity for this simplified case without pair
production, we obtain
\begin{eqnarray}
&&
\!\!\!\!\!\!\!\!\!\!\!\!\!\!\!\!\!\!\!\!\!\!
\partial_t\left [ \frac{(E-e_\pm A^0)\phi_\pm^2(\nabla_i S_\pm-e_\pm A^i)}{m_0+{\cal S}}\right ] +\sum_{j=1}^3 \nabla_j
\left [\frac {\phi_\pm^2 (\nabla_j S_\pm-e_\pm  A^j )  (\nabla_i S_\pm-e_\pm A^i ) }{m_0+{\cal S}} \right ]
\nonumber\\
&&
\!\!\!\!\!\!\!\!\!\!\!\!\!\!\!\!\!\!\!\!\!
=
-\frac{m}{m_0+{\cal S}}\sum_{j=1}^3 \nabla_j  p_{ij}^{(q)}
 - \phi_\pm^2\nabla_i {\cal S}
+\frac{1}{m_0+{\cal S}} \biggl \{ 
- (E-e_\pm A^0)\phi_\pm^2e_\pm F^{0i}
-\sum_{j=1}^3 \phi_\pm^2 (\nabla_j S_\pm-e_\pm  A^j )e_\pm F^{ij}
\biggr \}
\nonumber\\ 
&&
\!\!\!\!\!\!\!\!\!\!\!\!\!\!\!\!
+\frac{e_\pm\phi_\pm^2\nabla_i  A^0}{2(E-e_\pm A^0)(m_0+{\cal S})} 
\biggl  \{
(\nabla^2 \phi_\pm)/\phi_\pm-(\nabla S_\pm-e_\pm {\bf A} )^2 ]
-(m_0+{\cal S})^2+(E-e_\pm A^0)^2\biggr \}
\nonumber\\ 
&&
\!\!\!\!\!\!\!\!\!\!\!\!\!\!\!\!
-(E-e_\pm A^0)\phi_\pm^2(\nabla_i S_\pm-e_\pm A^i) 
\frac{\partial_t {\cal S}} {(m_0+{\cal S})^2}
-\sum_{j=1}^3 
\left [\frac {\phi_\pm^2 (\nabla_j S_\pm-e_\pm  A^j )  (\nabla_i S_\pm-e_\pm A^i ) }{m_0+{\cal S}} \right ]
\frac{\nabla_j {\cal S}} {(m_0+{\cal S})^2}. 
\end{eqnarray}
We can identity the fluid energy density $\epsilon_\pm$ as 
\begin{eqnarray}
\epsilon_\pm=(m_0+{\cal S})\phi_\pm^2,
\end{eqnarray}
as it corresponds to the fluid energy density for the fluid element at
rest.  The fluid element is characterized by a relativistic 4-velocity
$u^\mu$.  We can identify
\begin{eqnarray}
u_\pm^0&=&\frac{E-e_\pm A^0}{m_0+{\cal S}}
\nonumber\\
u_\pm^i&=&\frac{\nabla_i S_\pm-e_\pm A^i}{m_0+{\cal S}}, {\rm ~~~~for ~} i=1,2,3.
\end{eqnarray}
We can then write an equation of motion for $ \epsilon_\pm u_\pm^0 u_\pm^i$ \cite{Won10},
\begin{eqnarray}
\label{hyd}
  &&
\!\!\!\!\!\!\!\!\!\!\!\!\!\!\!\!\!\!\!\!\!\!
\partial_t ( \epsilon_\pm u_\pm^0 u_\pm^i)
  + \sum_{j=1}^3 \nabla_j  \epsilon_\pm u_\pm^i u_\pm^j
  +\frac{m}{m_0+{\cal S}}   \sum_{j=1}^3 \nabla_j p_{\pm ij}^{(q)}
  \nonumber\\
  && 
\!\!\!\!\!\!\!\!\!\!\!\!\!\!\!\!
=
  - \phi_\pm^2\nabla_i {\cal S}
  +\frac{1}{m_0+{\cal S}} \biggl \{ -
  (E-e_\pm A^0)\phi_\pm^2e_\pm F^{0i}
  -\sum_{j=1}^3 \phi_\pm^2 (\nabla_j S_\pm-e_\pm  A^j )e_\pm F^{ij}
  \biggr \}
  \nonumber\\ 
  &&
\!\!\!\!\!\!\!\!\!\!\!\!\!\!\!\!
  +\frac{e_\pm\phi_\pm^2 \nabla_i  A^0}{2(E-e_\pm A^0)(m_0+{\cal S})} 
  \biggl  \{
  (\nabla^2 \phi_\pm)/\phi_\pm-(\nabla S_\pm-e_\pm {\bf A} )^2 ]
  -(m_0+{\cal S})^2+(E-e_\pm A^0)^2\biggr \}
  \nonumber\\ 
&&
\!\!\!\!\!\!\!\!\!\!\!\!\!\!\!\!
  -(E-e_\pm A^0)\phi_\pm^2(\nabla_i S_\pm-e_\pm A^i) 
  \frac{\partial_t {\cal S}} {(m_0+{\cal S})^2}
  -\sum_{j=1}^3 
  \left [\frac {\phi_\pm^2 (\nabla_j S_\pm-e_\pm  A^j )  (\nabla_i S_\pm-e_\pm A^i ) }{m_0+{\cal S}} \right ]
  \frac{\nabla_j {\cal S}} {(m_0+{\cal S})^2},
  \nonumber\\ 
\end{eqnarray}
where $i,j=1,2,3.$ This is the Klein-Gordon equation for the particle
and antiparticle probability densities in hydrodynamical form.  The
first two terms on the left-hand side correspond to $\partial_\mu
T_\pm^{\mu i}$, with the energy momentum tensor of the probability
fluid $T_\pm^{\mu i}=\epsilon_\pm u_\pm^\mu u_\pm^i$, for
$\mu=0,1,2,3$.  The third term on the left-hand side is the quantum
stress tensor arising from the spatial variation of the amplitude of
the single-particle wave function \cite{Won76},
\begin{eqnarray}
p_{\pm\, ij}^{(q)} =-\frac{\hbar^2}{4m}\nabla^2\phi_\pm^2 \delta_{ij}
+\frac{\hbar^2}{m}\nabla_i \phi_\pm \nabla_j \phi_\pm.
\end{eqnarray}
Thus the dynamics of the probability fluid obeys an equation similar
to the hydrodynamical equation, with forces on fluid elements arising
from what one expects in classical considerations.  The additional
element is the presence of the quantum stress tensor $p_{ij}^{(q)}$
that is proportional to $\hbar^2$ and arises from the quantum nature
of the fluid.

\section{Relativistic  many-body system in 
the mean-field and no pair production approximation}

A many-body system in the time-dependent mean-field approximation
consists of a collection of independent particles moving in the
self-consistent mean-field generated by all other particles
\cite{Bon74,Won76,Won77}.  Each single-particle state $\psi_{a \nu}$
is characterized by a state label $a$, particle type $\nu$, energy
$e_\pm$, and occupation number $n_{a\nu}$.  Under the approximation
of no active pair
production by the mean field, the dynamics of the system is now described
by two distinct interacting fluids of particles and antiparticles.  We
consider the case in which the mean-field potential arises from a
scalar two-body interaction $v_s({\bf r}_1, {\bf r}_2)$ and a
time-like vector interaction $v_0({\bf r}_1, {\bf r}_2)$.  For
simplicity, we further neglect the last three terms on the right-hand
side of Eq.\ (\ref{hyd}) which represent higher-order relativistic
corrections.  The equation of motion for the energy density
$\epsilon_{a\nu}$ and velocity fields $u_{a\nu}^i$ for $i=1,2,3$ and
$\nu=\pm$, in the single particle state $a$ and particle type $\nu$,
is then
\begin{eqnarray}
\label{hyd24}
 \partial_t ( \epsilon_{a\nu} u_{a\nu}^0 u_{a\nu}^i)
  + \sum_{j=1}^3 \nabla_j  \epsilon_{a\nu} u_{a\nu}^i u_{a\nu}^j
  +\frac{m}{m_0+{\cal S}}   \sum_{j=1}^3 \nabla_j p_{ (a\nu)ij}^{(q)}
=
  - \phi_{a\nu}^2\nabla_i {\cal S}
  +\frac{E-e_{a\nu} A_\pm^0}{m_0+{\cal S}} 
  \phi_{a\nu}^2e_{a\nu} \frac{\partial A^0}{\partial x^i} ,
\end{eqnarray}
where, in the frame with the fluid element at rest,
\begin{eqnarray}
{\cal S}({\bf r},t) =\int d^3{\bf r}_2~ \rho({\bf r_2},t) v_s({\bf r},{\bf r}_2),
\end{eqnarray} 
\begin{eqnarray}
A^0({\bf r},t) =\int d^3{\bf r}_2~ \biggl \{ \rho_+({\bf r_2},t) e_+
+\rho_-({\bf r_2},t) e_-\biggr \}
 v_0({\bf r},{\bf r}_2),
\end{eqnarray}
\begin{eqnarray}
\rho_\nu=\sum_a n_{a\nu}\phi_{a\nu}^2, ~~ 
{\rm ~~and~~}
\rho=\rho_++\rho_-.
\end{eqnarray}
We consider a strongly interacting system in which the number of
particles and antiparticles are equal so that $\rho_+({\bf
  r_2})=\rho_-({\bf r_2})$ and $\rho_+({\bf r_2}) e_+ +\rho_-({\bf
  r_2}) e_-$ is zero.  Then the contribution from the second term on
the right-hand side of Eq.\ (\ref{hyd24}) is zero.  Multiplying
Eq.\ (\ref{hyd24}) by $n_{a\nu}$ and summing over $\{a,\nu\}$, we get
\begin{eqnarray} 
\!\!\!\!\!\!\!\!\!\!\!\!\!\!\!
\partial_t (\sum_{a\nu}n_{a\nu}\epsilon_{a\nu} u_{a\nu}^0 u_{a\nu}^i )
  + \sum_{j=1}^3 \nabla_j   (\sum_{a\nu}n_{a\nu}\epsilon_{a\nu} u_{a\nu}^i u_{a\nu}^j)
  +\frac{m}{m_0+{\cal S}}   \sum_{j=1}^3  \nabla_j ( \sum_{a\nu}n_{a\nu}p_{(a\nu) ij}^{(q)})
  + (\sum_{a\nu}n_{a\nu}\phi_{a\nu}^2) \nabla_i {\cal S} 
\nonumber\\
=  \sum_{a\nu}\epsilon_{a\nu} u_{a\nu}^0 u_{a\nu}^i [\partial_t n_{a\nu}]
\end{eqnarray}
We define the total energy density $\epsilon$ by
\begin{eqnarray}
\sum_{a\nu} n_{a\nu} \epsilon_{a\nu}=\epsilon,
\end{eqnarray}
and the average  4-velocity $u$ by 
\begin{eqnarray}
u = {\sum_{a\nu}  n_{a\nu} \epsilon_{a\nu} u_{a\nu}}
       /{\epsilon}.
\end{eqnarray}
We can introduce the thermal stress tensor $p_{ij}^{(t)}$ for
$\{i,j\}=1,2,3$ as the correlation of the deviations of the
single-particle velocity fields from the average
\begin{eqnarray}
\sum_{a\nu} n_{a\nu} \epsilon_{a\nu} (u_{a\nu}^i-u^i) (u_{a\nu}^j -u^j)
\equiv p_{ij}^{(t)}.
\end{eqnarray}
For the case with the suppression of pair production, we obtained the
Euler equation of motion for $\epsilon u^0 u^i$
\begin{eqnarray}
\partial_t (\epsilon u^0 u^i )
+ \sum_{j=1}^3 \left \{ \nabla_j \left  (\epsilon u^i u^j +p_{ij}^{(t)}+p_{ij}^{(v)}\right )
+\frac{m}{m_0+{\cal S}}  \nabla_j  p_{ ij}^{(q)} \right \}
= \sum_{a\nu}\epsilon_{a\nu} u_{a\nu}^0 u_{a\nu}^i [\partial_t n_{a\nu}] ,
\label{72}
\end{eqnarray}
where the total quantum stress tensor is
\begin{eqnarray}
p_{ ij}^{(q)} =-\frac{\hbar^2}{4m}\nabla^2\sum_{a\nu} n_{a\nu} \phi_{a\nu}^2 \delta_{ij}
+\frac{\hbar^2}{m}\sum_{a\nu}n_{a\nu}\nabla_i \phi_{a\nu} \nabla_j \phi_{a\nu},
\end{eqnarray}
and the pressure due to the interaction $p_{ij}^{(v)}$ is
\begin{eqnarray}
\frac{\partial}{\partial x^j} p_{ij}^{(v)}({\bf r},t)=n({\bf r},t)\nabla_i {\cal S}({\bf r},t)=
n({\bf r},t) \frac{\partial}{\partial x^j}
\int d^3 {\bf r}_2 n({\bf r}_2,t) v_s({\bf r},{\bf r}_2).
\end{eqnarray}

The Euler equation of motion (\ref{25}) for $\epsilon u^0 u^i$
contains the term that depends on the change of the occupation
probability $[\partial_t n_{a\nu}]$.  A finite system for which $n_{a\nu}$ is
quasi-stationary is said to be in thermal equilibrium, which can be
attained when $\partial_t n_{a\nu}$$\sim$ 0.  When thermal equilibrium is
reached, Eq.\ (\ref{72}) become the Euler equation in hydrodynamics.
The requirement of thermal equilibrium is one of the basis for
hydrodynamics.

The mean-field stress tensor $p_{ij}^{(v)}$ can also be given as
\begin{eqnarray}
 p_{ij}^{(v)}=
\left \{ n \frac{\partial (W^{(v)} n) }{\partial n}
- W^{(v)} n \right \}\delta_{ij},
\end{eqnarray}
where $W^{(v)}$ is the energy per particle arising from the mean-field
interaction.

The quantum stress tensor $p_{ij}^{(q)}$ depends on the amplitudes of
the wave functions while the thermal stress tensor
$p_{ij}^{(t)}$depends on the phases of the wave functions and the
deviation of the velocity fields from the mean velocities.  They can
take on different values, depending on the occupation numbers
$n_{a\nu}$ of the single-particle states that determine the degree of
thermal equilibrium of the system.  The quantum stress tensor is less
sensitive to the degree of thermalization as compared to the thermal
stress tensor.  In the time-dependent mean-field description, the
motion of each particle state can be individually followed
\cite{Bon74}.  The occupation numbers $n_{a\nu}$ of the
single-particle states will remain unchanged, if there are no
additional residual interaction between the single particles due to
residual interactions.  When particle residual interactions are
allowed as in the extended time-dependent mean-field approximation
\cite{Won76,Won77,Ayi80,Lac99}, the occupation numbers will change and
will approach an equilibrium distribution as time proceeds.

As the example of a finite potential in Section 3 indicates, the
quantum stress tensor depends on the wave functions which depends on
the geometry of the finite system.  In a system with anisotropic
shapes as in the overlapping dense region in a high-energy collision,
the initial geometrical shape will have important influences on the
quantum stress tensor in different directions and need to be explored
further.  If the mean-field potential is weakened by the occurrence of
a phase transition that alters the nature of the constituent
interactions in the new phase, then a non-isotropic hydrodynamical
flow of a pressurized medium will then occur.

\section{Summary and Discussions}

For dense systems with strongly interacting constituents, a reasonable
description of the system can be formulated in terms of constituents
moving in the strong mean field generated by all other particles.
Passage from the quantum mechanics to the hydrodynamics requires
further the assumption of a thermal equilibrium such that the
description of the occupation numbers of single-particle states in
terms of a temperature can be a reasonable concept.  It requires
furthermore the assumption of no active pair production by the mean field such that the
density flow and the momentum flow can be treated for particles and
antiparticles as different fluids.  From such an analysis, we find
that the probability density and the current of the system obey
hydrodynamical equations with the stress tensor arising from many
contributions.  There is the quantum stress tensor that arises from
quantum effects and wave functions, there is the thermal stress tensor
that arises from the deviation of the single-particle velocity fields
from the average velocity fields, and there is the mean-field stress
tensor that arises from the mean-field interactions.

The importance of the three different contributions depend on the
physical situations that are present in the system.  For low
temperature dynamics and finite systems with non-isotropic geometries
for which the quantum effects and mean field effects are important,
the dynamics of the strongly-coupled system will be influenced more by
the quantum stress tensor and the mean-field stress tensor than the
degree of thermalization.  On the other hand, for very high
temperatures for which the magnitude of the thermal stress tensor far
exceeds the strengths of the mean-field interactions and the quantum
pressure, the thermal stress tensor plays the dominant role and the
magnitude of the thermal stress tensor will depend sensitively on the
degree of thermalization of the system.  In between these limits, one
can envisage the transition from the quantum and mean field dominating
strongly-coupled regime to the thermal pressure dominating
weakly-coupled regime as temperature increases.

As both the Schr\"odinger equation and the Klein-Gordon equation for
bosons can be cast into a hydrodynamical form, one may inquire whether
the Dirac equation for fermions can also be written in hydrodynamical
form. It is well known that the Dirac equation can be reduced to a
Klein-Gordon equation, with additional terms involving the spin and
particle-antiparticle degrees of freedom.  For a Dirac particle in an
external field we have
\begin{eqnarray}
\{\gamma^\nu (i \partial_\pm - e A_\pm) - (m_0+{\cal S})\}
\psi = 0.
\end{eqnarray}
Upon multiplying this on the left by $\gamma^\nu (i \partial_\pm - e A_\pm) +
(m_0+{\cal S})$, we obtain
\begin{eqnarray}
\{ (i \partial_\pm - e A_\pm)^2 - (m_0+{\cal S})^2
-i {\bf \alpha}  \cdot  e {\bf E} + {\bf \sigma}\cdot e {\bf B}({\bf r}) 
- i [\gamma^\nu \partial_\pm {\cal S}]\} 
\psi = 0,
\end{eqnarray}
which is the Klein-Gordon equation with additional interactions.
Thus, the Dirac equation can be likewise cast into a hydrodynamical
form, following the procedures outlined in the present discussions.

What we have discussed is only a theoretical framework that exhibits
clearly the different sources of stress tensors.  To study
specifically the dynamics of the quark-gluon plasma for example, it
will be necessary to investigation the specific nature of different
constituents and their interactions in a case-by-case basis.
Nevertheless, the general roles played by the different components of
stress tensors can still be a useful reminder on the importance of the
quantum and mean-field stress tensors in the strongly-coupled regime,
at temperature just above the transition temperature $T_c$.

\vspace*{0.8cm}
\centerline {\bf Acknowledgment}

\vspace*{0.2cm} This research was supported in part by the Division of
Nuclear Physics, U.S. Department of Energy, under Contract No. DE-AC05-00OR22725.

\end{document}